\documentstyle[preprint,epsf,eqsecnum,aps]{revtex}
\input{epsf.sty}
\tightenlines

\begin{document}
\draft
\title{Aspects of the ground state of the $U=\infty$ Hubbard ladder
}
\author{Masanori Kohno}
\address{
Institute for Solid State Physics,\\
University of Tokyo, Roppongi, Minato-ku, Tokyo 106, Japan
}
\date{\today}
\maketitle
\begin{abstract}
We consider two aspects of the ground state of the $U=\infty$ Hubbard ladder:
ferromagnetism and the metal-insulator transition at quarter-filling. 
First, we present rigorous results for the $U=\infty$ Hubbard ladder in the limit of the large inter-chain hopping ($t_{\perp}/t\rightarrow\infty$).
In this limit, the total spin $S$ of the ground state is shown to be zero
for the electron density $n\le 0.5$ and its maximum ($S=S_{\rm max}$) 
for $n>0.5$. The charge gap at quarter-filling is $2t_{\perp}$.
We extend these results to finite $t_{\perp}/t$ by means of the density-matrix renormalization group method.
We estimate the phase boundaries with respect to spontaneous magnetization and the charge gap at quarter-filling for finite $t_{\perp}/t$. 
Applying the extended Aharonov-Bohm method, we give numerical evidence that the critical ratio $t_{\perp}/t$, above which the charge gap opens, is less than 0.001. 
Ferromagnetism in the $t$-$J$ ladder is briefly discussed. 
\end{abstract}

\narrowtext

\section{Introduction}
\label{Introduction}

Recently, a lot of attention has been given to the Hubbard or $t$-$J$ ladders. 
One of the reasons is that these models are believed to describe 
the essential features of various materials 
such as ${\rm (VO)}_2{\rm P}_2{\rm O}_7$\cite{VOPO} or ${\rm SrCu}_2{\rm O}_3$\cite{SrCuO}.
Another reason is that some properties of the ladder models are closer 
to planar models than to single chain models.
One such property is the enhancement of the $d_{x^2-y^2}$-like 
pairing correlation. 
In relation to high-$T_{\rm c}$ superconductivity\cite{HTC,Anderson}, 
this feature and the existence of the spin gap have been studied 
by various approaches such as bosonization\cite{Fabrizio,Khveshchenko,BalentsFisher}, 
projector Monte Carlo\cite{Kuroki}, exact diagonalization\cite{Troyer,HaywardED,Sano} 
and the density-matrix renormalization group (DMRG) method\cite{Noack,Hayward}.
\par
Another interesting aspect of the ladder models is the existence of the ferromagnetic ground state. 
Nagaoka's theorem\cite{Nagaoka,Thouless,Tasaki} holds 
for the ladders and the planar models, although not for the single chains.
Hence, ferromagnetic ground states may exist 
in the strong-coupling regime near half-filling 
for the ladders and the planar models, but not for the single chains.
For two dimensions, many workers have investigated the ferromagnetic ground state at $U=\infty$ for finite hole-density
\cite{MTakaFerro,KuboFerro,Riera,Yedidia,Zhang,Putikka,Kusakabe1,Liang}, 
in order to clarify the origin of itinerant ferromagnetism 
from the electron-correlation view-point.
In spite of such attempts, results are inconclusive, 
because of the lack of efficient methods.
On the other hand, for the ladders, there is an efficient method, 
i.e. the density-matrix renormalization group (DMRG) method proposed by White\cite{White}. Liang and Pang\cite{Liang} applied this method to the $U=\infty$ Hubbard ladder for $t_{\perp}/t=1$, 
and obtained some indications of ferromagnetism.
As for the (two-leg) ladder, the DMRG calculation in ref.\cite{Liang} suggests that the fully-polarized ferromagnetic state is one of the ground states for $\delta<0.22$ ($\delta$ : hole density) and that the ground state is a spin-singlet for $\delta\buildrel > \over \sim 0.4$.
They have tried to extend these results to two dimensions by investigating multi-leg ladders.
In this paper, we extend their results to various values of $t_{\perp}$ and $J$ for the $t$-$J$ ladder, in order to understand the ground-state properties of the Hubbard or $t$-$J$ ladders in the strong-coupling regime.
\par
Another interesting feature of strongly correlated electron systems is the metal-insulator (MI) transition. 
For single chains and planar models, there have been a lot of works on the MI transition\cite{LiebWu,Mott1D,Furukawa,Imada,Assaad,Kohno}. 
However, for ladder models, there are relatively few. One of the characteristic features of ladder models is the band structure. In the weak-coupling regime ($U\simeq 0$), the low-energy physics for $n\le 1$ is effectively described by the bonding band, if $t_{\perp}/t$ is large enough. Thus, the MI transition in this parameter regime is essentially the same as that of the single chains\cite{BalentsFisher}.
This argument may be true for $t_{\perp}\gg U$. 
A natural question arising here is what happens in the opposite limit, i.e. $U\gg t_{\perp}$. In order to answer this question, we consider the MI transition in the $U=\infty$ Hubbard ladder for simplicity.
\par
In the present paper, we mainly consider the $U=\infty$ Hubbard ladder (or $t$-ladder), which is defined as the $t$-$J$ ladder at $J=J_{\perp}=0$. Later, in Sec.\ref{DMRGFerro}, we briefly discuss ferromagnetism of the $t$-$J$ ladder. The Hamiltonian of the $t$-$J$ ladder is defined as follows :
\begin{eqnarray}
   {\cal H}_{tJ} &=& {\cal H}_t+{\cal H}_J, \\
   {\cal H}_t &=& -t\sum_{i \sigma}
(\tilde{c}^{1 \dagger}_{i\sigma}\tilde{c}^{1}_{i+1\sigma}+\tilde{c}^{2 \dagger}_{i\sigma}\tilde{c}^{2}_{i+1\sigma}+{\mbox{h.c.}})\nonumber\\
&&-t_{\perp}\sum_{i \sigma}
(\tilde{c}^{1 \dagger}_{i\sigma}\tilde{c}^{2}_{i\sigma}+{\mbox{h.c.}}), \nonumber\\
   {\cal H}_J &=& J\sum_{i}(\mbox{\boldmath $S^{1}_i\cdot S^{1}_{i+1}$}-\frac{1}{4}n^{1}_in^{1}_{i+1}+\mbox{\boldmath $S^{2}_i\cdot S^{2}_{i+1}$}-\frac{1}{4}n^{2}_in^{2}_{i+1})\nonumber\\
&&+J_{\perp}\sum_{i}(\mbox{\boldmath $S^{1}_i\cdot S^{2}_i$}-\frac{1}{4}n^{1}_in^{2}_i),\nonumber
\end{eqnarray}
where $\tilde{c}^{\alpha\dagger}_{i\sigma}$ denotes a creation operator of
an electron at site $i$ with spin $\sigma
(\sigma=\uparrow,\downarrow)$ in the $\alpha$-th chain ($\alpha=1,2$) with the constraint that no site is
doubly occupied, i.e. $\tilde{c}^{\alpha\dagger}_{i\sigma}\equiv c^{\alpha\dagger}_{i\sigma}(1-n^{\alpha}_{i -\sigma})$. 
The number operator $n^{\alpha}_{i\sigma}$ is
defined as $n^{\alpha}_{i\sigma}\equiv c^{\alpha\dagger}_{i\sigma}c^{\alpha}_{i\sigma}$, using
the standard electron creation operator $c^{\alpha\dagger}_{i\sigma}$. The
spin operator at site $i$ in the $\alpha$-th chain is defined as $\mbox{\boldmath
$S$}^{\alpha}_i\equiv\frac{1}{2}\sum_{\beta\gamma}c^{\alpha\dagger}_{i\beta}\mbox{\boldmath
$\sigma_{\beta\gamma}$}c^{\alpha}_{i\gamma}$, where $\mbox{\boldmath
$\sigma$ }$ is the vector of Pauli matrices. 
The number of electrons, the number of sites and the number of rungs are 
denoted by $N_e$, $N_s$ and $L(=N_s/2)$, respectively. 
The electron density $n$ and the hole density $\delta$ are defined 
as $n\equiv N_e/N_s$ and $\delta\equiv 1-n$, respectively. The maximum value of the total spin $S$ is denoted by $S_{\rm max}(=N_e/2)$.
\par
This paper is organized as follows:
In Sec \ref{Proof}, we present rigorous results on the ground state 
of the $t$-ladder in the limit $t_{\perp}/t\rightarrow\infty$. 
In Sec \ref{DMRGFerro}, we present numerical results on ferromagnetism in the $t$-ladder and the $t$-$J$ ladder.
In Sec \ref{MItrans}, we discuss the metal-insulator transition at quarter-filling for the $t$-ladder. Section \ref{Summary} is a summary.

\section{$t$-ladder in the limit $t_{\perp}/t\rightarrow\infty$}
\label{Proof}
In this section, we prove the following statements:
\begin{enumerate}
\item The ground state of the $t$-ladder 
in the limit $t_{\perp}/t\rightarrow\infty$ for $n\le 0.5$ is a spin-singlet ($S=0$) and is unique in finite-size clusters with an even number of electrons with open boundary conditions.
\label{Theorem1}
\item The ground state of the $t$-ladder 
in the limit $t_{\perp}/t\rightarrow\infty$ for $n>0.5$ has the maximum total spin ($S=S_{\rm max}$) and is unique 
up to the trivial $(N_e+1)$-fold degeneracy  
in finite-size clusters with open boundary conditions.
\label{Theorem2}
\end{enumerate}
\par
Before investigating the above properties, we consider the $t$-ladder at $t=0$.
The ground states of this model can be written in the following form:
\begin{equation}
|\Phi\rangle=\bigotimes_{i=1}^L|\alpha\rangle_{i},
\label{unperturb}
\end{equation}
where $|\alpha\rangle_{i}$'s correspond to either of states (i)-(iii) defined in table \ref{BOD} for $n\le 0.5$, and (ii)-(vii) for $n>0.5$.
The degeneracy of the ground states, the energy $E$ and the chemical potential $\mu$($\equiv \partial E/\partial N_e$) are summarized in table \ref{t0}. The charge gap $\Delta_c$ at quarter-filling ($n=0.5$) is $2t_{\perp}$.
\par
Next, we consider the $t$-ladder in the limit $t/t_{\perp}\rightarrow 0$.
Let us consider the cases $n\le 0.5$ and $n>0.5$, separately.

\subsection{$n\le 0.5$}
Up to order $t$, the effective Hamiltonian is
\begin{equation}
{\cal H}^{\rm eff}_{t} = -t\sum_{i \sigma}
(\tilde{b}^{\dagger}_{i\sigma}\tilde{b}_{i+1\sigma}+{\mbox{h.c.}}),
\end{equation}
where $\tilde{b}^{\dagger}_{i\sigma}$ denotes a creation operator of
an electron in the bonding band at rung $i$ with spin $\sigma
(\sigma=\uparrow,\downarrow)$ with the constraint that no rung is
doubly occupied, or
$\tilde{b}^{\dagger}_{i\sigma}\equiv b^{\dagger}_{i\sigma}(1-n^1_i-n^2_i)$.
Here, the creation operator $b^{\dagger}_{i\sigma}$ is defined as $b^{\dagger}_{i\sigma}\equiv(\tilde{c}^{1\dagger}_{i\sigma}+\tilde{c}^{2\dagger}_{i\sigma})/\sqrt 2$. This Hamiltonian is equivalent to the $U=\infty$ Hubbard chain. Thus, the ground states are degenerate with respect to the spin degrees of freedom. The charge part of the ground-state wavefunction is simply that of the spinless fermion model on a chain.
\par
Next, we consider the effective Hamiltonian of order $t^2/t_{\perp}$.
Here, we define the local Hamiltonian ${\cal H}^{\rm loc}_{i,i+1\sigma}$:
\begin{equation}
{\cal H}^{\rm loc}_{i,i+1\sigma} = -t
(\tilde{c}^{1 \dagger}_{i\sigma}\tilde{c}^{1}_{i+1\sigma}+\tilde{c}^{2 \dagger}_{i\sigma}\tilde{c}^{2}_{i+1\sigma}+{\mbox{h.c.}}).
\end{equation}
Letting this local Hamiltonian operate on $|\alpha\rangle_{i}\otimes|\beta\rangle_{i+1}$, we obtain the following relations:
\begin{eqnarray}
{\cal H}^{\rm loc}_{i,i+1\sigma}|B\sigma\rangle_{i}\otimes|B\sigma\rangle_{i+1}&=&0,\nonumber\\
{\cal H}^{\rm loc}_{i,i+1\sigma}|B-\sigma\rangle_{i}\otimes|B\sigma\rangle_{i+1}&=&\frac{\sigma t}{\sqrt 2}|S\rangle_{i}\otimes|00\rangle_{i+1},\nonumber\\
{\cal H}^{\rm loc}_{i,i+1\sigma}|B\sigma\rangle_{i}\otimes|B-\sigma\rangle_{i+1}&=&-\frac{\sigma t}{\sqrt 2}|00\rangle_{i}\otimes|S\rangle_{i+1},\nonumber\\
{\cal H}^{\rm loc}_{i,i+1\sigma}|S\rangle_{i}\otimes|00\rangle_{i+1}&=&\frac{\sigma t}{\sqrt 2}(|B-\sigma\rangle_{i}\otimes|B\sigma\rangle_{i+1}-|A-\sigma\rangle_{i}\otimes|A\sigma\rangle_{i+1}),\nonumber\\
{\cal H}^{\rm loc}_{i,i+1\sigma}|00\rangle_{i}\otimes|S\rangle_{i+1}&=&-\frac{\sigma t}{\sqrt 2}(|B\sigma\rangle_{i}\otimes|B-\sigma\rangle_{i+1}-|A\sigma\rangle_{i}\otimes|A-\sigma\rangle_{i+1}),\nonumber
\end{eqnarray}
where $|S\rangle_{i}$ and $|A\sigma\rangle_{i}$ are defined as $\frac{1}{\sqrt 2}(\tilde{c}^{1\dagger}_{i\uparrow}\tilde{c}^{2\dagger}_{i\downarrow}-\tilde{c}^{1\dagger}_{i\downarrow}\tilde{c}^{2\dagger}_{i\uparrow})|00\rangle_{i}$ and  $\frac{1}{\sqrt 2}(\tilde{c}^{1\dagger}_{i\sigma}-\tilde{c}^{2\dagger}_{i\sigma})|00\rangle_{i}$, respectively. Thus, the second-order perturbation energy $E_2$ is obtained as in table \ref{2nd}.
It is also shown that the following relations are satisfied:
\begin{eqnarray}
n^{B}_{i\uparrow}n^{B}_{i+1\downarrow}|B\uparrow\rangle_{i}\otimes|B\downarrow\rangle_{i+1}&=&|B\uparrow\rangle_{i}\otimes|B\downarrow\rangle_{i+1},\nonumber\\
S^{B-}_{i}S^{B+}_{i+1}|B\uparrow\rangle_{i}\otimes|B\downarrow\rangle_{i+1}&=&|B\downarrow\rangle_{i}\otimes|B\uparrow\rangle_{i+1},\nonumber
\end{eqnarray}
where $S^{B+}_{i}\equiv\tilde{b}^{\dagger}_{i\uparrow}\tilde{b}_{i\downarrow}$ and $n^{\rm B}_{i\sigma}\equiv \tilde{b}^{\dagger}_{i\sigma}\tilde{b}_{i\sigma}$.
Thus, the effective Hamiltonian ${\cal H}^{\rm eff}$, up to order $t^2/t_{\perp}$, is written as follows:
\begin{eqnarray}
\label{Th1eq}
{\cal H}^{\rm eff} &=& {\cal H}^{\rm eff}_{t}+{\cal H}^{\rm eff(1)}_{J}+{\cal H}^{\rm eff(2)}_{J},\\ \nonumber
{\cal H}^{\rm eff(1)}_{J} &=& J_{\rm eff}\sum_{i}
(\mbox{\boldmath $S^{B}_i\cdot S^{B}_{i+1}$}-\frac{1}{4}n^{B}_in^{B}_{i+1}),\\ \nonumber
{\cal H}^{\rm eff(2)}_{J} &=& \frac{J_{\rm eff}}{4}\sum_{i \sigma}
(\tilde{b}^{\dagger}_{i-1-\sigma}\tilde{b}^{\dagger}_{i\sigma}\tilde{b}_{i-\sigma}\tilde{b}_{i+1\sigma}-\tilde{b}^{\dagger}_{i-1\sigma}n^{B}_{i-\sigma}\tilde{b}_{i+1\sigma}+{\mbox{h.c.}}),
\end{eqnarray}
where $\mbox{\boldmath $S^{B}_i$}$ is the spin operator in the bonding band at rung $i$ and $J_{\rm eff}=t^2/t_{\perp}$.
Hence, it is shown that the effective Hamiltonian, up to order $t^2/t_{\perp}$ for $n\le 0.5$, has the same form as that of the $U\rightarrow\infty$ Hubbard chain, up to order $t^2/U$ for $n\le 1$\cite{MTakaU,Hirsh,Gros,Ogata}.
As a result, the ground-state properties of the $t$-ladder in the limit $t_{\perp}/t\rightarrow\infty$ for $n\le 0.5$ are the same as those of the $U\rightarrow\infty$ Hubbard chain for $n\le 1$\cite{Ogata}. This leads to a spin-singlet ground state by the Lieb-Mattis theorem\cite{LiebMattes}.

\subsection{$n>0.5$}
In this subsection, we consider the case $n>0.5$.
The unperturbed ground states are written in the form of eq.(\ref{unperturb}).
The matrix elements of the local Hamiltonian eq.(\ref{localH}) are summarized in table \ref{matrix}.
\begin{equation}
\label{localH}
{\cal H}^{\rm loc}_{i,i+1} = -t\sum_{\sigma}
(\tilde{c}^{1 \dagger}_{i\sigma}\tilde{c}^{1}_{i+1\sigma}+\tilde{c}^{2 \dagger}_{i\sigma}\tilde{c}^{2}_{i+1\sigma}+{\mbox{h.c.}}).
\end{equation}
Considering the matrix elements in table \ref{matrix}, it is shown that the state $|B-\sigma\rangle_{i}\otimes|\sigma\sigma\rangle_{i+1}$ can reach the state $|\sigma\sigma\rangle_{i}\otimes|B-\sigma\rangle_{i+1}$ after successive multiplication by the local Hamiltonian ${\cal H}^{\rm loc}_{i,i+1}$ as follows: $|B-\sigma\rangle_{i}\otimes|\sigma\sigma\rangle_{i+1}\rightarrow|\sigma-\sigma\rangle_{i}\otimes|B\sigma\rangle_{i+1}\rightarrow|B\sigma\rangle_{i}\otimes|\sigma-\sigma\rangle_{i+1}\rightarrow|\sigma\sigma\rangle_{i}\otimes|B-\sigma\rangle_{i+1}$.
Using this property, we can show that the following processes are possible by 
applying the local Hamiltonians ${\cal H}^{\rm loc}_{i-1,i}$ and ${\cal H}^{\rm loc}_{i,i+1}$ successively.
\begin{eqnarray}
|B\sigma\rangle_{i-1}\otimes|\sigma\sigma\rangle_{i}\otimes|\sigma-\sigma\rangle_{i+1}&\leftrightarrow&|B\sigma\rangle_{i-1}\otimes|\sigma-\sigma\rangle_{i}\otimes|\sigma\sigma\rangle_{i+1},\nonumber\\
|B\sigma\rangle_{i-1}\otimes|\sigma\sigma\rangle_{i}\otimes|-\sigma\sigma\rangle_{i+1}&\leftrightarrow&|B\sigma\rangle_{i-1}\otimes|-\sigma\sigma\rangle_{i}\otimes|\sigma\sigma\rangle_{i+1},\nonumber\\
|B\sigma\rangle_{i-1}\otimes|-\sigma-\sigma\rangle_{i}\otimes|-\sigma\sigma\rangle_{i+1}&\leftrightarrow&|B\sigma\rangle_{i-1}\otimes|-\sigma\sigma\rangle_{i}\otimes|-\sigma-\sigma\rangle_{i+1},\nonumber\\
|B\sigma\rangle_{i-1}\otimes|-\sigma-\sigma\rangle_{i}\otimes|\sigma-\sigma\rangle_{i+1}&\leftrightarrow&|B\sigma\rangle_{i-1}\otimes|\sigma-\sigma\rangle_{i}\otimes|-\sigma-\sigma\rangle_{i+1},\nonumber\\
|B\sigma\rangle_{i-1}\otimes|\sigma\sigma\rangle_{i}\otimes|-\sigma-\sigma\rangle_{i+1}&\leftrightarrow&|B\sigma\rangle_{i-1}\otimes|-\sigma-\sigma\rangle_{i}\otimes|\sigma\sigma\rangle_{i+1},\nonumber\\
|B\sigma\rangle_{i-1}\otimes|\sigma-\sigma\rangle_{i}\otimes|-\sigma\sigma\rangle_{i+1}&\leftrightarrow&|B\sigma\rangle_{i-1}\otimes|-\sigma\sigma\rangle_{i}\otimes|\sigma-\sigma\rangle_{i+1}.\nonumber
\end{eqnarray}
Thus, $|\uparrow\uparrow\rangle$, $|\uparrow\downarrow\rangle$, $|\downarrow\uparrow\rangle$ and  $|\downarrow\downarrow\rangle$ can have their positions changed after successive multiplication by the Hamiltonian (${\cal H}_t$), if there exists at least one $|B\uparrow\rangle$ or $|B\downarrow\rangle$.
Furthermore, the following processes are also possible:
\begin{eqnarray}
|\sigma\sigma\rangle_{i-1}\otimes|B\sigma\rangle_{i}\otimes|B-\sigma\rangle_{i+1}&\leftrightarrow&|\sigma\sigma\rangle_{i-1}\otimes|B-\sigma\rangle_{i}\otimes|B\sigma\rangle_{i+1},\nonumber\\
|\sigma-\sigma\rangle_{i-1}\otimes|B\sigma\rangle_{i}\otimes|B-\sigma\rangle_{i+1}&\leftrightarrow&|\sigma-\sigma\rangle_{i-1}\otimes|B-\sigma\rangle_{i}\otimes|B\sigma\rangle_{i+1}.\nonumber
\end{eqnarray}
Thus, $|B\uparrow\rangle$ and $|B\downarrow\rangle$ can have their positions changed, if there exists at least one $|\uparrow\uparrow\rangle$, $|\uparrow\downarrow\rangle$, $|\downarrow\uparrow\rangle$ or $|\downarrow\downarrow\rangle$.
As a result, any unperturbed ground state in the form of eq.(\ref{unperturb}) can be reached after successive multiplication by the Hamiltonian (${\cal H}_t$) for $0.5<n<1$ in the subspace of fixed total $S^z$ and fixed electron number. This property is usually called connectivity.
Combined this property and the fact that the matrix elements of the Hamiltonian in this representation are all positive (table \ref{matrix}), the Perron-Frobenius theorem ensures that the state of the largest eigenvalue is unique and the wavefunction is positive (nodeless) in this representation in each subspace.
\par
Next, we consider the ground-state wavefunction in the subspace of the maximum total $S^z$ ($S^z=S_{\rm max}$).
The Hamiltonian in this subspace has the same form 
as that of the spinless fermion model on a chain.
Thus, the wavefunction of the largest eigenvalue in this subspace is nothing but that of the spinless fermion model ($|\Psi_{\rm SF}\rangle$).
Applying the spin-lowering operator $S^-$ to $|\Psi_{\rm SF}\rangle$, we obtain the eigenstates of various total $S^z$'s. These states have all positive (nodeless) wavefunctions in the present representation. Thus, these states have finite overlap with the states of the largest eigenvalue in the corresponding subspaces. 
By use of the gauge transformation $\tilde{c}^{\alpha}_{i}\rightarrow(-1)^i\tilde{c}^{\alpha}_{i}, \alpha=1,2$,
the sign of the hopping amplitude $t$ can be changed, i.e. $t\rightarrow -t$, with spin operators unchanged. As a result, it is shown that the ground state of the $t$-ladder in the limit $t_{\perp}/t\rightarrow\infty$ for $0.5<n<1$ has the maximum total spin ($S=S_{\rm max}$) and is unique up to the trivial $(N_e+1)$-fold degeneracy.  
\subsection{Remarks}
Here, we give some remarks on the above theorems.
\begin{enumerate}
\item Part of theorem \ref{Theorem1} can be extended to higher dimensions.
The effective Hamiltonian of double-layer $t$-models up to order $t^2/t_{\perp}$ for $n\le 0.5$ has the same form as that of the single-layer Hubbard models up to order $t^2/U$ for $n\le 1$\cite{MTakaU,Hirsh,Gros}. 
Thus, the ground-state properties of double-layer $t$-models in the limit $t_{\perp}/t\rightarrow\infty$ for $n\le 0.5$ are the same as those of the single-layer $U\rightarrow\infty$ Hubbard models for $n\le 1$\cite{Lieb}.
\item The proof of theorem \ref{Theorem2} is mathematically similar to that of Kubo's theorem\cite{Kubo}. However, the physical situation is different.
In Kubo's theorem, the limit of the strong Hund-coupling is taken, i.e. $H_{\rm Hund}\equiv -J_{\rm H}\sum_i\mbox{\boldmath $S^1_i\cdot S^2_{i}$}$, 
$J_{\rm H}\rightarrow \infty$. Furthermore, almost degenerate bands are assumed.
On the other hand, in theorem \ref{Theorem2},
we do not assume explicit ferromagnetic couplings. In contrast to Kubo's case, the limit of the large band splitting is taken in the present case. 
The extension of Kubo's theorem to $n\le 0.5$ shows that the ground state is also ferromagnetic\cite{Kusakabe2}, which is contrasted with theorem \ref{Theorem1}.
The proof of theorem \ref{Theorem2} is mathematically similar to that of ref.\cite{Kondo} for the one-dimensional Kondo-lattice model, too.
\item The restriction on the boundary condition can be relaxed such that the Hubbard chain has a unique spin-singlet ground state for theorem \ref{Theorem1} and that the spinless fermion model on a chain has all positive matrix elements in the site representation for theorem \ref{Theorem2}. For example, theorem \ref{Theorem2} can be extended to the case of periodic boundary conditions with an odd number of electrons 
and an even number of rungs.
Nagaoka's theorem is recovered for the one-hole case\cite{Nagaoka,Thouless,Tasaki}
\end{enumerate}

\section{Ferromagnetism}
\label{DMRGFerro}
In this section, we present the numerical results 
on the $t$-ladder for finite $t_{\perp}/t$ and the $t$-$J$ ladder for small $J$ and $J_{\perp}$ obtained by the DMRG method (finite-size algorithm)\cite{White}. The DMRG calculation has been performed with open boundary conditions.
\par
As a test of the DMRG calculation, we compare the ground-state energy obtained by the DMRG method with that of the exact diagonalization method. 
In Fig.\ref{diag}, the agreement of the data obtained by these two methods is good. The maximum error is about 0.01\%. 
Next, we consider the truncation error, i.e. the error due to small $m$, where $m$ is the number of states kept in the superblock\cite{White}.
The difference between $m=50$ and $m=100$ is very small as shown in Fig.\ref{m50m100}, suggesting that $m=50$ is sufficient. 
(See also Fig.\ref{SzJ0}.) Thus, we mainly report the results of $m=50$ hereafter.
\par
Let us consider ferromagnetism of the $t$-ladder.
In Fig. \ref{dEJ0}, we show the energy difference $\Delta E_{\rm F}$ between the ground-state energy in the subspace of $S^z=0$ and that of $S^z=S_{\rm max}$ as a function of filling. The data in various-size clusters are scaled to a single line, indicating that the finite-size error is small. (See also Fig.\ref{ChemJ0}.) 
From this plot, we find the region FF where the fully-polarized ferromagnetic state is one of the ground states. The phase boundary $n_{c_1}$ between FF and non-FF is estimated as shown in Fig. \ref{PDJ0}. 
At $t_{\perp}/t=1$, the result in ref.\cite{Liang} ($n_{c_1}\simeq0.78$) is recovered. Qualitatively, the region FF becomes larger as $t_{\perp}/t$ increases. The phase boundary $n_{c_1}$ gets closer to 0.5 as $t_{\perp}/t$ increases. This is consistent with the rigorous results in Sec.\ref{Proof}.
\par
Next, we consider the phase boundary $n_{c_2}$ between SS and non-SS, where SS is defined as the region in which the ground state is a spin-singlet.
Figure \ref{SzJ0} shows the ground-state energy as a function of the total $S^z$ at $t_{\perp}/t=1$ for $n=0.5625$, $0.625$ and $0.6875$.
The ground state is a spin-singlet for $n=0.5625$ and not for $n=0.625$ and $0.6875$. Hence, the phase boundary $n_{c_2}$ is estimated as $n_{c_2}=0.59\pm0.03$, which is consistent with ref.\cite{Liang}($n_{c_2}\simeq 0.6$).
In this way, the phase boundary $n_{c_2}$ is estimated for various $t_{\perp}/t$ as shown in Fig.\ref{PDJ0}. The region PF shrinks as $t_{\perp}/t$ increases, where PF is defined as the region which is neither FF nor SS. This is also consistent with the rigorous results in Sec.\ref{Proof}.
\par
Here, we present the numerical results on the $t$-$J$ ladder for small $J$ and $J_{\perp}$.
For simplicity, we choose $t_{\perp}=t$ and $J_{\perp}=J$, and set $t=1$ as the energy unit.
Figure \ref{dEJ} shows the $n$-dependence of $\Delta E_{\rm F}$ at $J=0.00, 0.05, 0.07, 0.10$ and $0.15$. As shown in this figure, $\Delta E_{\rm F}$ becomes large near half-filling ($n=1$) due to antiferromagnetic correlation, and $\Delta E_{\rm F}$ seems to be the smallest near $n\simeq0.8$ for $n>0.5$. Figure \ref{dEJ} also indicates that $J=0.05$ is enough to destroy the region FF.
\par
Next, we consider the stability of PF against $J$. The ground-state energy as a function of the total $S^z$ near half-filling for  $J=0.05, 0.07, 0.10$ and $0.15$ is shown in Fig. \ref{SzJ}. This figure suggests that FF is surrounded by PF in the phase diagram of the $t$-$J$ ladder for finite $\delta$.

\section{Metal-insulator transition at quarter-filling}
\label{MItrans}
Before discussing the metal-insulator transition, we consider the charge gap $\Delta_c$ at quarter-filling ($n=0.5$).
Figure \ref{ChemJ0} shows the $n$-dependence of the chemical potential $\mu$ ($\equiv\partial E/\partial  N_e$).
The chemical potential $\mu$ in a finite-size cluster is defined as 
\begin{equation}
\mu(\bar n)\equiv \frac{E(n_1)-E(n_2)}{(n_1-n_2)N_s},
\end{equation}
where $E(n_i)$ denotes the ground-state energy at filling $n_i$, $i=1,2$, and $\bar n\equiv (n_1+n_2)/2$.
We took $(n_2-n_1)N_s=2$. The charge gap $\Delta_c$ is defined as $\Delta_c\equiv\mu(n_c+0)-\mu(n_c-0)$, where $n_c$ is the critical electron-density ($n_c=0.5$ in the present case). It is expected that the charge gap opens at quarter-filling, if $t_{\perp}/t$ is large enough (Sec.\ref{Proof}). 
Actually, for large values of $t_{\perp}/t$, the charge gap seems to open 
as shown in Fig.\ref{ChemJ0}(a). For smaller values of $t_{\perp}/t$, we cannot determine whether the charge gap opens from Fig.\ref{ChemJ0}. Thus, we extrapolate the charge gap in a finite-size cluster [eq.(\ref{gapeq})] as $a+b/L$, using the data for $L=12-24$, and estimate the charge gap as shown in Fig. \ref{Chargegap}.
\begin{equation}
\label{gapeq}
\Delta_c(N_e=N_s/2)\equiv \frac{E(N_e=N_s/2+2)+E(N_e=N_s/2-2)-2E(N_e=N_s/2)}{2}.
\end{equation}
\par
There are some possibilities for the critical value $t_{\perp c}/t$ above which the charge gap opens.
One of the possibilities is that the critical value $t_{\perp c}/t$ is zero and that the gap is exponentially small in the limit $t_{\perp}/t\rightarrow 0$ as in the case of the Hubbard chain in the limit $U\rightarrow 0$\cite{LiebWu}. 
In order to determine the critical value $t_{\perp c}/t$, we adopt the extended Aharonov-Bohm (AB) method proposed by Kusakabe and Aoki\cite{KusakabeAB}. 
In the framework of this method, we investigate the extended spectral flow by introducing a Peierls phase as
\begin{equation}
\tilde{c}^{\alpha \dagger}_{i\sigma}\tilde{c}^{\alpha}_{i+1\sigma}\rightarrow\exp({\rm i}\frac{2\pi\Phi}{L\Phi_0})\tilde{c}^{\alpha \dagger}_{i\sigma}\tilde{c}^{\alpha}_{i+1\sigma}.
\end{equation}
It is expected that the period of the spectral flow reduces to $L\Phi_0/M$ if  $M$-particle bound states are formed in the ground state. For example, $M=1$ for a metallic state, and $M=2$ for a BCS state.
We apply this method to the $t$-ladder with a very small value of $t_{\perp}/t$ ($t_{\perp}/t=0.001$).
As shown in Fig.\ref{ABflux1}, the spectral flow at quarter-filling has the minimum extended AB period, i.e. $\Phi=\Phi_0$,  
suggesting that the ground state is an $L$-particle bound state or, in this case, an insulator\cite{AritaAB}. 
This behavior is contrasted with the case off quarter-filling. 
For example, at $n=1/3$, the extended AB period is larger than $\Phi_0$ as shown in Fig.\ref{ABflux2}. 
This implies that the ground state at quarter-filling is an insulator already for $t_{\perp}/t=0.001$. 
It is plausible to consider that the critical value $t_{\perp c}/t$ may be zero. A possible scenario may be the following: The perturbation of the small $t_{\perp}$-term produces the relevant Umklapp process which leads to an insulator, at the same time as the degeneracy with respect to the spin degrees of freedom is removed.
The numerical results presented above are quite different from the case in the weak-coupling limit ($U\rightarrow+0$). In the weak-coupling limit, it is shown by bosonization that $t_{\perp c}/t$ is one\cite{BalentsFisher}. Thus, it is expected that $t_{\perp c}/t$ decreases from 1 to 0 as the interaction $U$ increases from 0 to $\infty$.
\par
Now, let us consider the metal-insulator (MI) transition at quarter-filling.
As discussed in Sec. \ref{Proof}, in the limit $t_{\perp}/t\rightarrow \infty$, the MI transition for $n\rightarrow 0.5-0$ is effectively described by the equivalent model to the $U\rightarrow\infty$ Hubbard chain, 
and the MI transition for $n\rightarrow0.5+0$ is described by the spinless fermion model on a chain.
It is interesting to compare these features 
with those in the weak-coupling regime.
In the weak-coupling regime ($U\rightarrow+0$), the charge gap is also expected at quarter-filling for $t_{\perp}/t>1$ because of the relevant Umklapp process\cite{BalentsFisher}.
This MI transition is understood as the Mott transition which is described by the $U\rightarrow+0$ Hubbard model on a chain written in terms of the bonding-band operators.
In both weak-coupling [$U\ll t_{\perp}(>t)$] and strong-coupling [$U\gg t_{\perp} (\gg t$)] regimes, as $n\rightarrow 0.5-0$, the MI transition is described by single chain effective Hubbard Hamiltonians. However, the value of the charge gap will have different energy scales.
In the weak-coupling regime, the value of the charge gap would be determined mainly by $U$. On the other hand, in the strong-coupling regime, it would be determined mainly by $t_{\perp}$. This feature is similar to the two types of the MI transition for transition-metal compounds\cite{MottCharge}, i.e. the Mott-Hubbard type and the charge-transfer type.

\section{Summary}
\label{Summary}
In summary, two aspects of the ground state of the $U=\infty$ Hubbard ladder are investigated. One is ferromagnetism, the other is the metal-insulator (MI) transition. 
In the limit $t_{\perp}/t\rightarrow\infty$, it is rigorously shown that the ground state is a spin-singlet for $n\le 0.5$ and that the total spin is maximum for $0.5<n<1$.
For finite $t_{\perp}/t$, we have estimated the phase boundaries, with respect to spontaneous magnetization, by the density-matrix renormalization group method.
It is numerically shown that the region FF becomes larger and spreads down to quarter-filling as $t_{\perp}/t$ increases, which is consistent with the rigorous results presented in Sec.\ref{Proof}. The rigorous results ($t_{\perp}/t\rightarrow\infty$) and the numerical results for finite $t_{\perp}/t$ support one another and confirm that the ground state can be ferromagnetic for the $U=\infty$ Hubbard ladder with finite hole-density. The numerical results for the $t$-$J$ ladder suggest that FF is surrounded by PF for finite $\delta$ in the small $J$ regime. We have also estimated the value of the charge gap at quarter-filling ($n=0.5$). Applying the extended Aharonov-Bohm method, we have obtained numerical evidence that the critical value $t_{\perp c}/t$, above which the charge gap opens, is less than 0.001. This is quite different from that of the weak-coupling limit ($U\rightarrow+0$) ($t_{\perp c}/t=1$)\cite{BalentsFisher}. 

\narrowtext
\acknowledgments
The author would like to thank M. Takahashi, M. Ogata, K. Kusakabe and F.V. Kusmartsev for helpful discussions and useful comments. 
The author also thanks D. Lidsky for reading of the manuscript. 
The exact diagonalization program is partly based on the subroutine package "TITPACK Ver.2" coded by H. Nishimori.
Part of the calculations were performed on the Fujitsu VPP500 at Institute for Solid State Physics, Univ. of Tokyo.

\begin{table}
\caption{Basis set.}
\label{BOD}
\begin{center}
\begin{tabular}{cccc}
 &Symbol&Definition&Energy\\ \hline
(i)&$|00\rangle_{i}$&${\rm vacuum}$&0\\
(ii)&$|B\uparrow\rangle_{i}$&$\frac{1}{\sqrt 2}(\tilde{c}^{1\dagger}_{i\uparrow}+\tilde{c}^{2\dagger}_{i\uparrow})|00\rangle_{i}$&$-t_{\perp}$\\
(iii)&$|B\downarrow\rangle_{i}$&$\frac{1}{\sqrt 2}(\tilde{c}^{1\dagger}_{i\downarrow}+\tilde{c}^{2\dagger}_{i\downarrow})|00\rangle_{i}$&$-t_{\perp}$\\
(iv)&$|\uparrow\uparrow\rangle_{i}$&$\tilde{c}^{1\dagger}_{i\uparrow}\tilde{c}^{2\dagger}_{i\uparrow}|00\rangle_{i}$&0\\
(v)&$|\downarrow\downarrow\rangle_{i}$&$\tilde{c}^{1\dagger}_{i\downarrow}\tilde{c}^{2\dagger}_{i\downarrow}|00\rangle_{i}$&0\\
(vi)&$|\uparrow\downarrow\rangle_{i}$&$\tilde{c}^{1\dagger}_{i\uparrow}\tilde{c}^{2\dagger}_{i\downarrow}|00\rangle_{i}$&0\\
(vii)&$|\downarrow\uparrow\rangle_{i}$&$\tilde{c}^{1\dagger}_{i\downarrow}\tilde{c}^{2\dagger}_{i\uparrow}|00\rangle_{i}$&0
\end{tabular}
\end{center}
\end{table}

\begin{table}
\caption{Ground states at $t=0$.}
\label{t0}
\begin{center}
\begin{tabular}{ccc}
 &$n<0.5$&$0.5<n<1$\\ \hline
Degeneracy &$_{L}{\rm C}_{N_e}\times 2^{N_e}$&$_{L}{\rm C}_{N_e-L}\times 2^{N_e}$\\
Energy $E$&$-t_{\perp}\times N_e$&$-t_{\perp}\times (2L-N_e)$\\
Chemical Potential $\mu$&$-t_{\perp}$&$t_{\perp}$\\
\end{tabular}
\end{center}
\end{table}

\begin{table}
\caption{Second-order perturbation energy $E_2$.}
\label{2nd}
\begin{center}
\begin{tabular}{ccc}
$E_2$ between $|\alpha\rangle$ and $|\beta\rangle$ & $|\alpha\rangle$&$|\beta\rangle$\\ \hline \nonumber
0&$|B\sigma\rangle_{i}\otimes|B\sigma\rangle_{i+1}$&$|B\sigma\rangle_{i}\otimes|B\sigma\rangle_{i+1}$\\ \nonumber
$-t^2/(2t_{\perp})$&$|B\sigma\rangle_{i}\otimes|B-\sigma\rangle_{i+1}$&$|B\sigma\rangle_{i}\otimes|B-\sigma\rangle_{i+1}$\\ \nonumber
$t^2/(2t_{\perp})$&$|B\sigma\rangle_{i}\otimes|B-\sigma\rangle_{i+1}$&$|B-\sigma\rangle_{i}\otimes|B\sigma\rangle_{i+1}$\\ \nonumber
$t^2/(4t_{\perp})$&$|00\rangle_{i-1}\otimes|B\sigma\rangle_{i}\otimes|B-\sigma\rangle_{i+1}$&$|B\sigma\rangle_{i}\otimes|B-\sigma\rangle_{i}\otimes|00\rangle_{i+1}$\\ \nonumber
$-t^2/(4t_{\perp})$&$|00\rangle_{i-1}\otimes|B\sigma\rangle_{i}\otimes|B-\sigma\rangle_{i+1}$&$|B-\sigma\rangle_{i}\otimes|B\sigma\rangle_{i}\otimes|00\rangle_{i+1}$
\end{tabular}
\end{center}
\end{table}

\begin{table}
\caption{Matrix elements.}
\label{matrix}
\begin{center}
\begin{tabular}{cccc}
$\langle\alpha|{\cal H}^{\rm loc}_{i,i+1}|\beta\rangle$&$|\alpha\rangle$&$|\beta\rangle$\\ \hline \nonumber
$t$&$|B\sigma\rangle_{i}\otimes|\sigma\sigma\rangle_{i+1}$&$|\sigma\sigma\rangle_{i}\otimes|B\sigma\rangle_{i+1}$\\ \hline \nonumber
&$|B\sigma\rangle_{i}\otimes|\sigma-\sigma\rangle_{i+1}$&$|\sigma-\sigma\rangle_{i}\otimes|B\sigma\rangle_{i+1}$\\ \nonumber
&$|B\sigma\rangle_{i}\otimes|\sigma-\sigma\rangle_{i+1}$&$|\sigma\sigma\rangle_{i}\otimes|B-\sigma\rangle_{i+1}$\\ \nonumber
$t/2$&$|B\sigma\rangle_{i}\otimes|-\sigma\sigma\rangle_{i+1}$&$|-\sigma\sigma\rangle_{i}\otimes|B\sigma\rangle_{i+1}$\\ \nonumber
&$|B\sigma\rangle_{i}\otimes|-\sigma\sigma\rangle_{i+1}$&$|\sigma\sigma\rangle_{i}\otimes|B-\sigma\rangle_{i+1}$\\ \nonumber
&$|B\sigma\rangle_{i}\otimes|-\sigma-\sigma\rangle_{i+1}$&$|-\sigma\sigma\rangle_{i}\otimes|B-\sigma\rangle_{i+1}$\\ \nonumber
&$|B\sigma\rangle_{i}\otimes|-\sigma-\sigma\rangle_{i+1}$&$|\sigma-\sigma\rangle_{i}\otimes|B-\sigma\rangle_{i+1}$\\ \hline \nonumber
0& otherwise
\end{tabular}
\end{center}
\end{table}

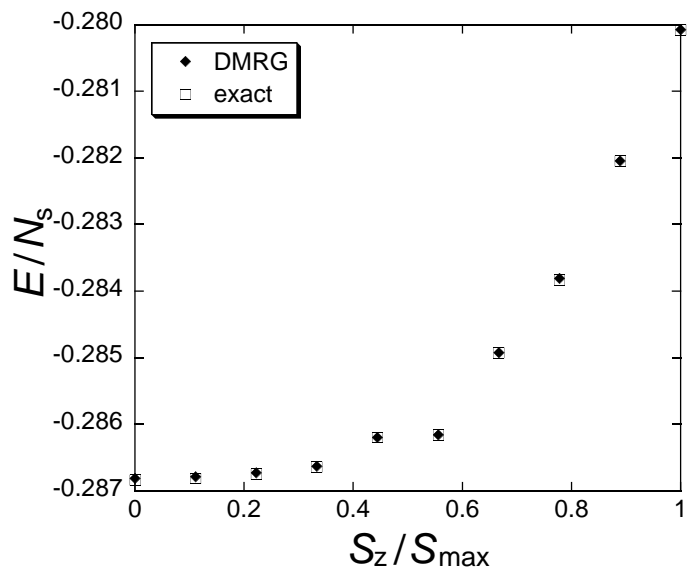
\begin{figure}
\caption{Energy per site as a function of the total $S^z$, measured from that of  $S^z=S_{\rm max}$, for the $t$-$J$ ladder at $J=J_{\perp}=0.05$ and $t=t_{\perp}=1$ in a $10\times 2$-site cluster with 18 electrons. The solid diamonds and open squares correspond to the data obtained by the DMRG method and the exact diagonalization method, respectively. For the DMRG method, we took $m=50$ ($m$ : the number of states kept in the superblock\protect\cite{White}) and repeated 2-3 sweeps for convergence.}
\label{diag}
\end{figure}

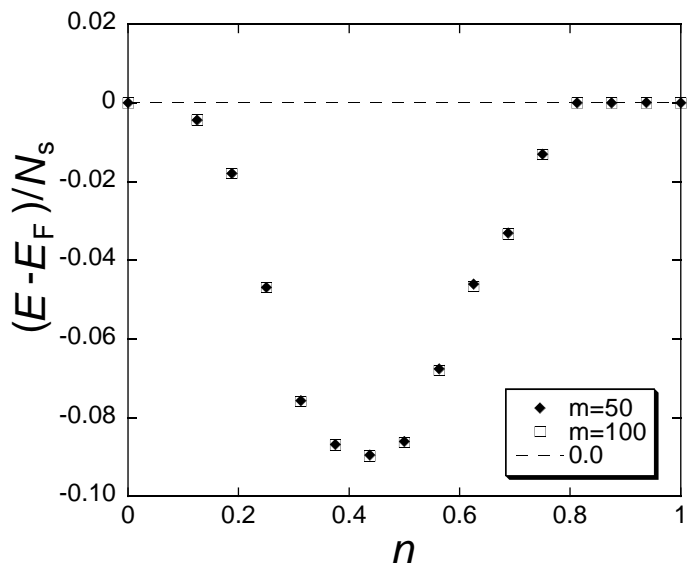
\begin{figure}
\caption{Energy difference $\Delta E_{\rm F}$ per site between the ground-state energy in the subspace of $S^z=0$ and that of $S^z=S_{\rm max}$ as a function of filling for the $t$-ladder at $t_{\perp}/t=1$. The solid diamonds and open squares correspond to the data of $m=50$ and $m=100$, respectively. $t=1$.}
\label{m50m100}
\end{figure}

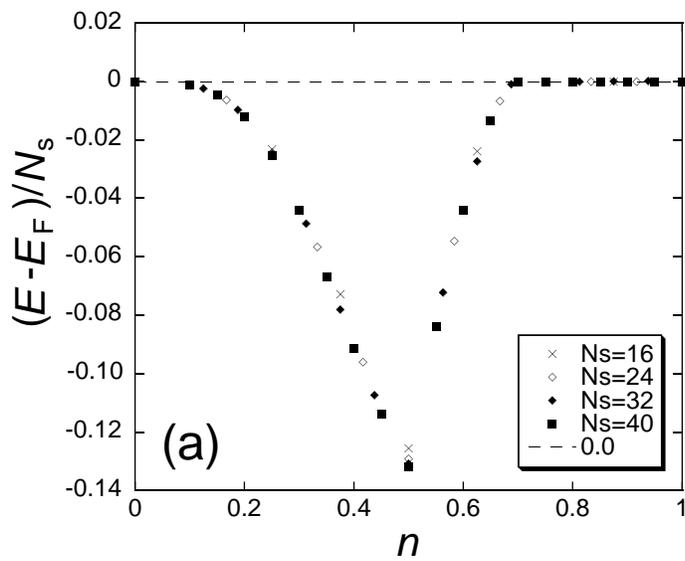
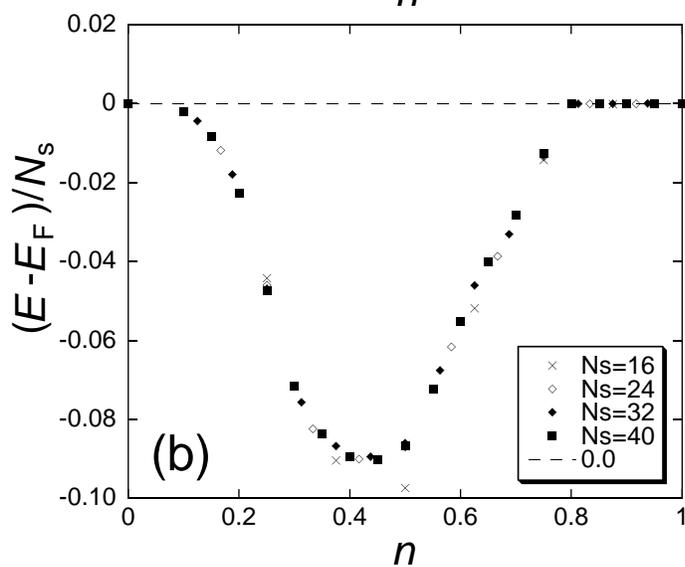
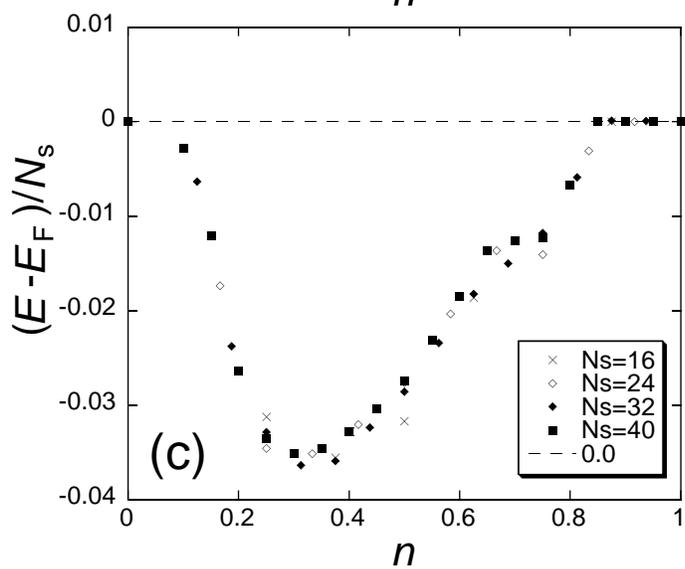
\begin{figure}
\caption{Energy difference $\Delta E_{\rm F}$ per site as a function of filling for the $t$-ladder at $t_{\perp}/t=$2.5 [(a)], 1.0 [(b)] and 0.5 [(c)]. $t=1$.}
\label{dEJ0}
\end{figure}

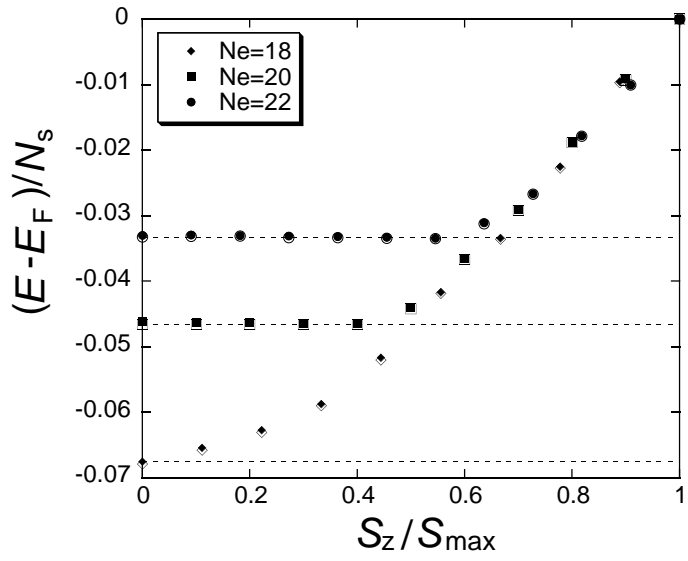
\begin{figure}
\caption{Energy per site as a function of the total $S^z$, measured from that of  $S^z=S_{\rm max}$, for the $t$-ladder at $t=t_{\perp}=1$ in a $16\times 2$-site cluster with 22, 20 and 18 electrons starting from above. The solid and open symbols denote the data for $m=50$ and $m=100$, respectively.}
\label{SzJ0}
\end{figure}

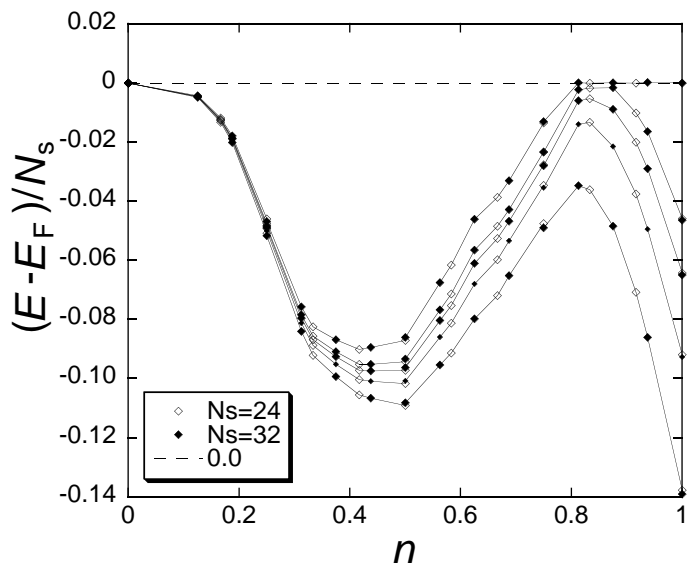
\begin{figure}
\caption{Energy difference $\Delta E_{\rm F}$ per site as a function of filling for the $t$-$J$ ladder in $12\times 2$-site and $16\times 2$-site clusters at $J/t=J_{\perp}/t=0.00, 0.05, 0.07, 0.10$ and $0.15$ starting from above. $t=t_{\perp}=1$.}
\label{dEJ}
\end{figure}

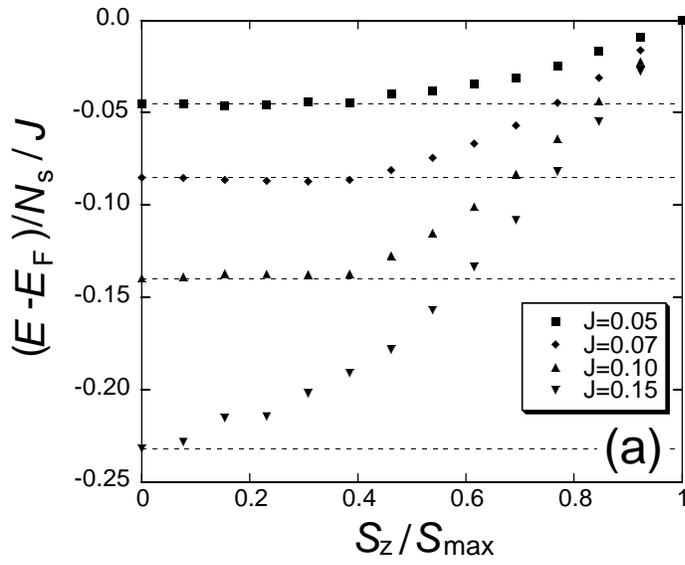
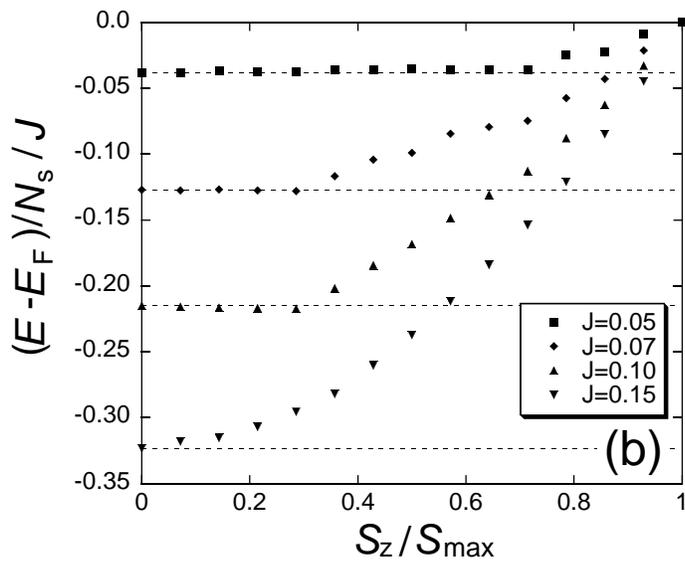
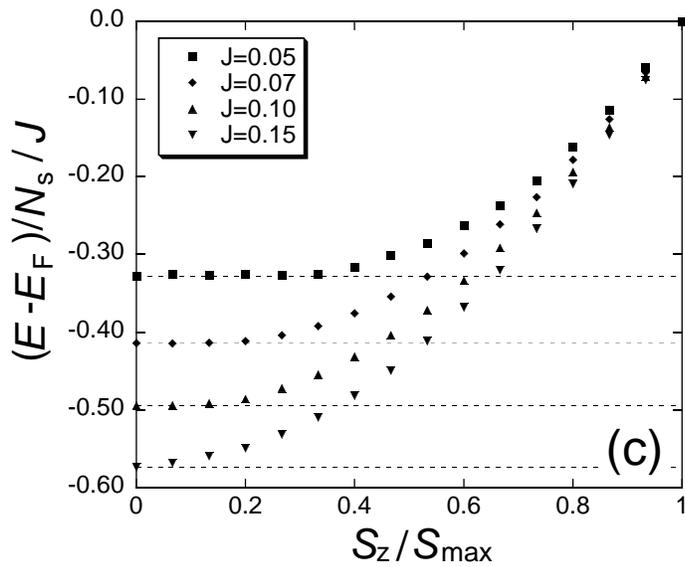
\begin{figure}
\caption{Energy per site as a function of the total $S^z$, measured from that of  $S^z=S_{\rm max}$, for the $t$-$J$ ladder at $t=t_{\perp}=1$ in a $16\times 2$-site cluster with 26 [(a)], 28 [(b)] and 30 [(c)] electrons at $J/t=J_{\perp}/t=$0.05, 0.07, 0.10 and 0.15 starting from above. The dotted lines correspond to the energy of $S^z=0$. The data are normalized by $J$.}
\label{SzJ}
\end{figure}

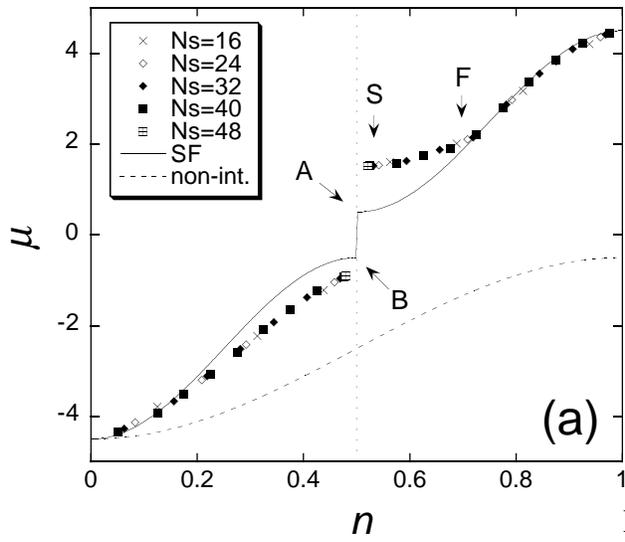
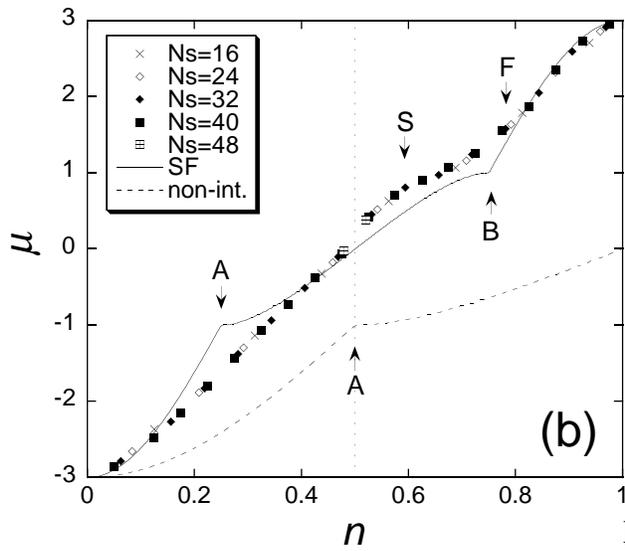
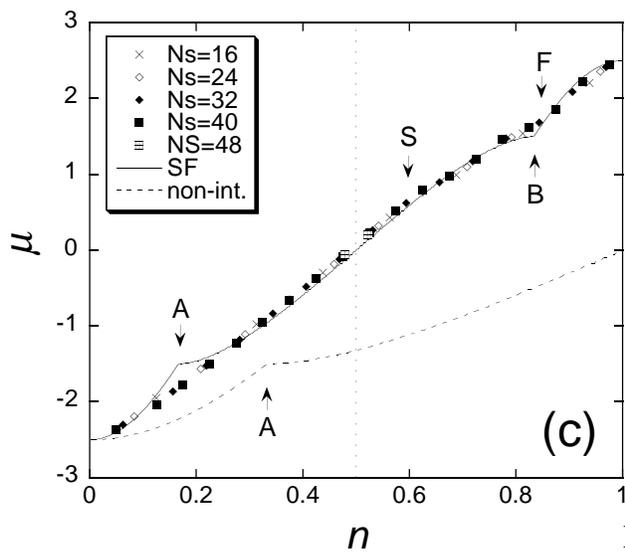
\begin{figure}
\caption{Chemical potential $\mu$ as a function of filling at $t_{\perp}/t=$2.5 [(a)], 1.0 [(b)] and 0.5 [(c)]. For comparison, the chemical potential $\mu$ for the non-interacting case (dotted line) and that of the spinless fermion model (solid line) are shown ($N_s=160\times 2$). The points A and B correspond to the anomalies due to the band bottom of the anti-bonding band and the band top of the bonding band, respectively. The points F and S correspond to the phase boundaries $n_{c_1}$ and $n_{c_2}$, respectively, which are estimated by the DMRG method (Sec.\protect\ref{DMRGFerro}). $t=1$.}
\label{ChemJ0}
\end{figure}

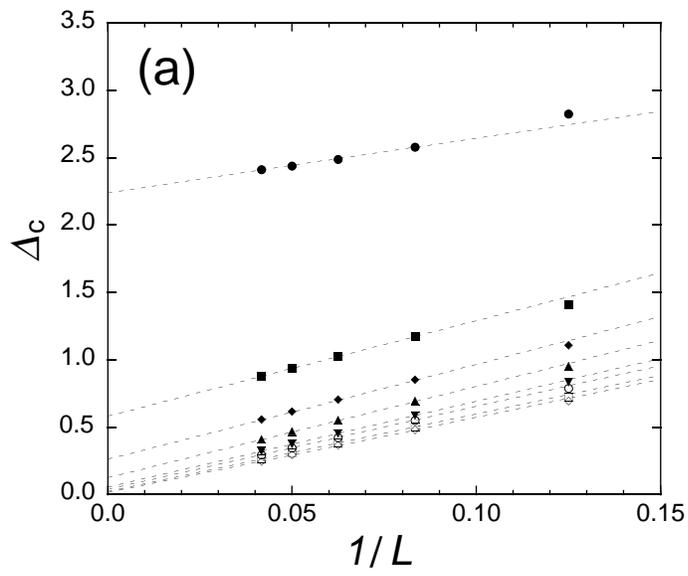
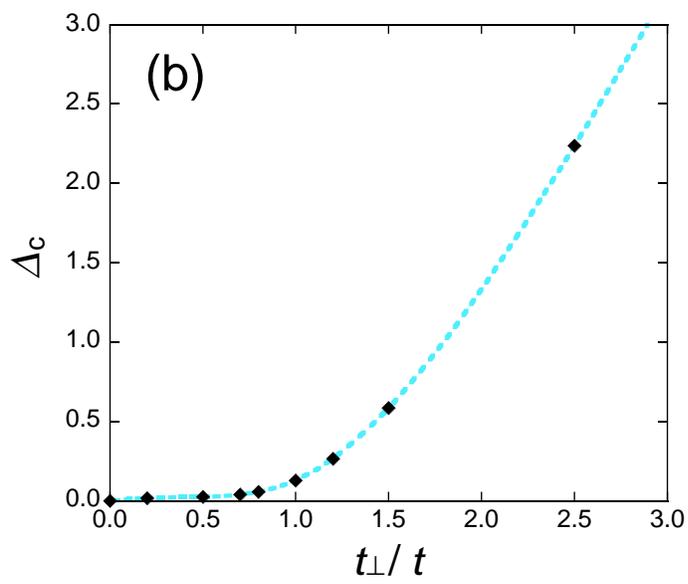
\begin{figure}
\caption{(a) Size dependence of the charge gap $\Delta_c$ for the $t$-ladder at $t_{\perp}/t=$2.5, 1.5, 1.2, 1.0, 0.8, 0.7, 0.5 and 0.2 starting from above. 
(b) The charge gap $\Delta_c$ as a function of $t_{\perp}/t$. The bold line is a guide to the eye. $t=1$.}
\label{Chargegap}
\end{figure}

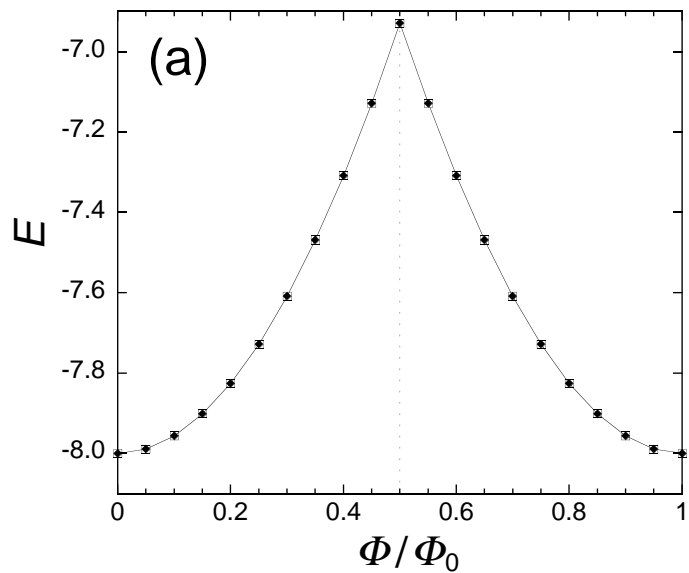
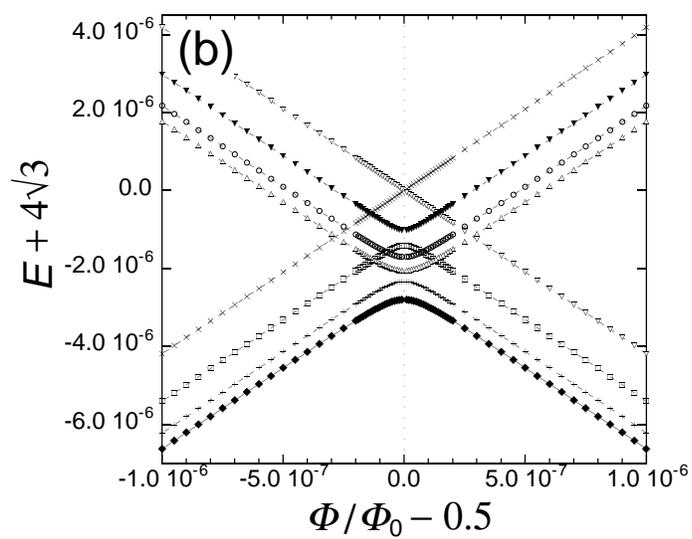
\begin{figure}
\caption{Spectral flow of the $t$-ladder for $t_{\perp}/t=0.001$ in a 12-site cluster with 6 electrons for $0 \le \Phi/\Phi_0\le 1$ [(a)], 
and the blow-up region for $\Phi/\Phi_0\simeq 0.5$ [(b)]. 
The solid diamonds correspond to the spectral flow of the ground state. 
The data are obtained by the exact diagonalization method. $t=1$.}
\label{ABflux1}
\end{figure}

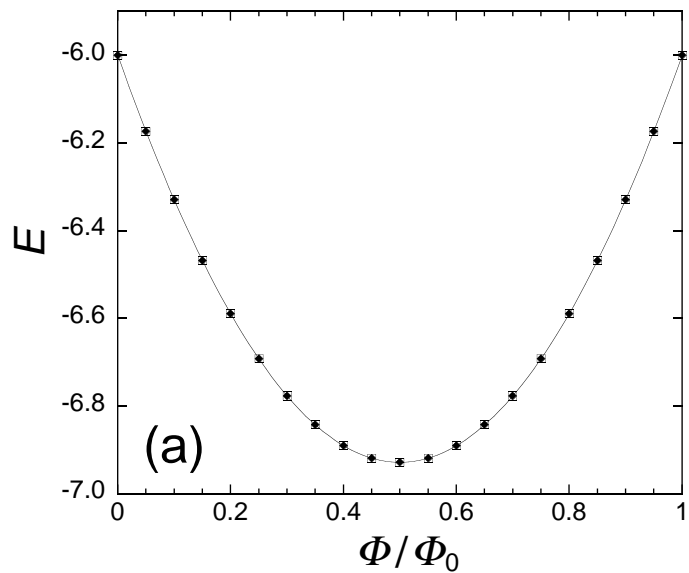
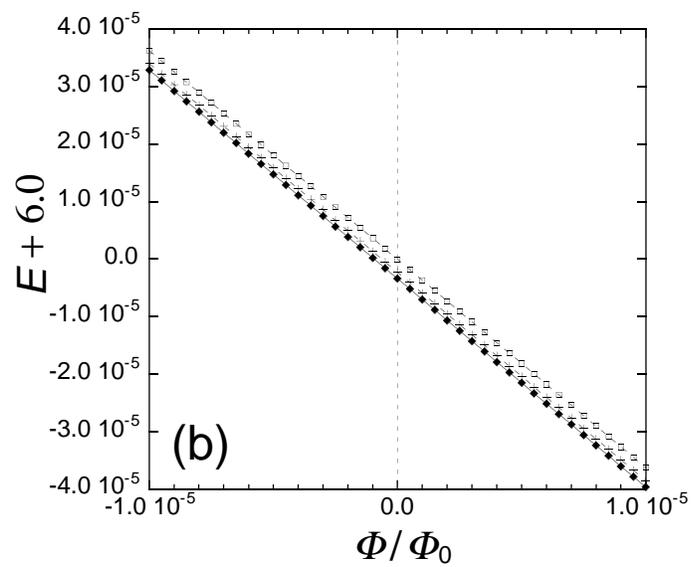
\begin{figure}
\caption{Spectral flow of the $t$-ladder for $t_{\perp}/t=0.001$ in a 12-site cluster with 4 electrons for $0 \le \Phi/\Phi_0\le 1$ [(a)], 
and the blow-up region for $\Phi/\Phi_0\simeq 0$ [(b)]. 
The solid diamonds correspond to the spectral flow of the ground state. 
The data are obtained by the exact diagonalization method. $t=1$.}
\label{ABflux2}
\end{figure}

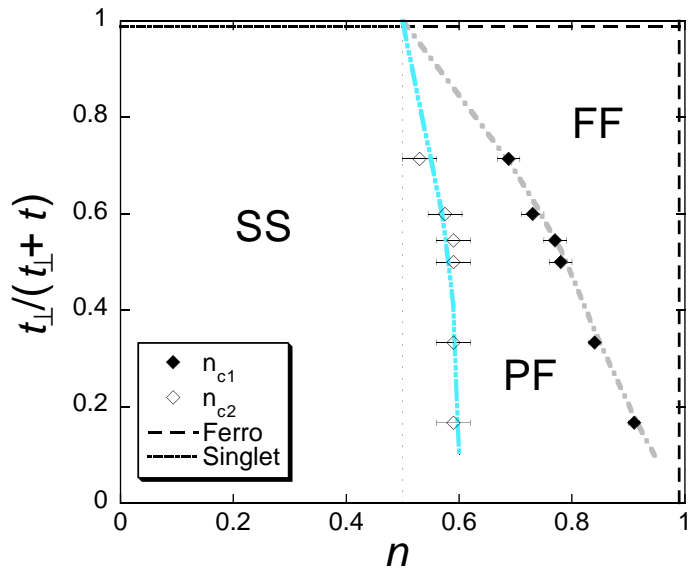
\begin{figure}
\caption{Phase diagram of the $U=\infty$ Hubbard ladder with respect to spontaneous magnetization. The dashed lines denote the region where the ground state is rigorously shown to have the maximum total spin $S$ ($S=S_{\rm max}$) (Theorem \protect\ref{Theorem2} and Nagaoka's theorem).  The dotted line in the limit $t_{\perp}/t\rightarrow\infty$ corresponds to the region where the ground state is shown to be a spin-singlet (Theorem \protect\ref{Theorem1}). The solid and open diamonds correspond to the phase boundaries $n_{c_1}$ and $n_{c_2}$, estimated by the DMRG method. The bold lines are guides to the eye. At $t_{\perp}/t=0$, the ground states are degenerate with respect to the spin degrees of freedom, because the model reduces to the decoupled $U=\infty$ Hubbard chains. At quarter-filling, the charge gap is expected (Sec.\protect\ref{MItrans}).}
\label{PDJ0}
\end{figure}

\newpage
\epsfxsize=3.5in\epsfbox{tmodelFIG.dir/szDMRGexact.epsi}
Fig.\ref{diag}\\
\vspace{2mm}\\
\epsfxsize=3.5in\epsfbox{tmodelFIG.dir/diffL16m50m100.epsi}
\vspace{2mm}\\
Fig.\ref{m50m100}\\
\epsfxsize=3.5in\epsfbox{tmodelFIG.dir/Fig3a.epsi}
Fig.\ref{dEJ0}(a)\\
\epsfxsize=3.5in\epsfbox{tmodelFIG.dir/Fig3b.epsi}
Fig.\ref{dEJ0}(b)
\epsfxsize=3.5in\epsfbox{tmodelFIG.dir/Fig3c.epsi}
Fig.\ref{dEJ0}(c)
\epsfxsize=3.5in\epsfbox{tmodelFIG.dir/szL16J0m50m100.epsi}
\vspace{2mm}\\
Fig.\ref{SzJ0}\\
\vspace{2mm}\\
\epsfxsize=3.5in\epsfbox{tmodelFIG.dir/diffJtp10.epsi}
\vspace{2mm}\\
Fig.\ref{dEJ}\\
\epsfxsize=3.5in\epsfbox{tmodelFIG.dir/Fig6a.epsi}
Fig.\ref{SzJ}(a)
\vspace{2mm}\\
\epsfxsize=3.5in\epsfbox{tmodelFIG.dir/Fig6b.epsi}
Fig.\ref{SzJ}(b)
\vspace{2mm}\\
\epsfxsize=3.5in\epsfbox{tmodelFIG.dir/Fig6c.epsi}
Fig.\ref{SzJ}(c)
\epsfxsize=3.2in\epsfbox{tmodelFIG.dir/Fig7a.epsi}
Fig.\ref{ChemJ0}(a)
\vspace{2mm}\\
\epsfxsize=3.2in\epsfbox{tmodelFIG.dir/Fig7b.epsi}
Fig.\ref{ChemJ0}(b)
\vspace{2mm}\\
\epsfxsize=3.2in\epsfbox{tmodelFIG.dir/Fig7c.epsi}
Fig.\ref{ChemJ0}(c)
\vspace{2mm}
\epsfxsize=3.5in\epsfbox{tmodelFIG.dir/Fig8a.epsi}
\vspace{2mm}\\
Fig.\ref{Chargegap}(a)\\
\vspace{2mm}\\
\epsfxsize=3.5in\epsfbox{tmodelFIG.dir/Fig8b.epsi}
\vspace{2mm}\\
Fig.\ref{Chargegap}(b)\\
\vspace{2mm}\\
\epsfxsize=3.5in\epsfbox{tmodelFIG.dir/Fig9a.epsi}
\vspace{2mm}\\
Fig.\ref{ABflux1}(a)\\
\vspace{2mm}\\
\epsfxsize=3.5in\epsfbox{tmodelFIG.dir/Fig9b.epsi}
\vspace{2mm}\\
Fig.\ref{ABflux1}(b)\\
\vspace{2mm}\\
\epsfxsize=3.5in\epsfbox{tmodelFIG.dir/Fig10a.epsi}
\vspace{2mm}\\
Fig.\ref{ABflux2}(a)\\
\vspace{2mm}\\
\epsfxsize=3.5in\epsfbox{tmodelFIG.dir/Fig10b.epsi}
\vspace{2mm}\\
Fig.\ref{ABflux2}(b)\\
\vspace{2mm}\\
\epsfxsize=3.5in\epsfbox{tmodelFIG.dir/PhD.epsi}
\vspace{2mm}\\
Fig.\ref{PDJ0}

\end{document}